\documentclass[12pt]{article}
\usepackage{graphicx}
\usepackage{epsfig}
\usepackage{latexsym}
\usepackage{latexsym}
\usepackage{amssymb}
\usepackage{cite}
\newcommand{\reef}[1]{(\ref{#1})}

\DeclareSymbolFont{AMSb}{U}{msb}{m}{n}
\DeclareMathSymbol{\IN}{\mathbin}{AMSb}{"4E}
\DeclareMathSymbol{\IZ}{\mathbin}{AMSb}{"5A}
\DeclareMathSymbol{\IR}{\mathbin}{AMSb}{"52}
\DeclareMathSymbol{\Q}{\mathbin}{AMSb}{"51}
\DeclareMathSymbol{\II}{\mathbin}{AMSb}{"49}
\DeclareMathSymbol{\IC}{\mathbin}{AMSb}{"43}
\DeclareMathSymbol{\IP}{\mathbin}{AMSb}{"50}
\DeclareMathSymbol{\IH}{\mathbin}{AMSb}{"48}
\DeclareMathSymbol\IA{\mathalpha}{AMSb}{"41}
\DeclareMathSymbol\IS{\mathalpha}{AMSb}{"53}

\def\Q{{\cal Q}}

\begin{document}

\begin{flushright}
USC-03-06\\
NSF-KITP-03-99\\
DCPT--03/51
\end{flushright}
{\flushleft\vskip-1.35cm\vbox{\psfig{figure=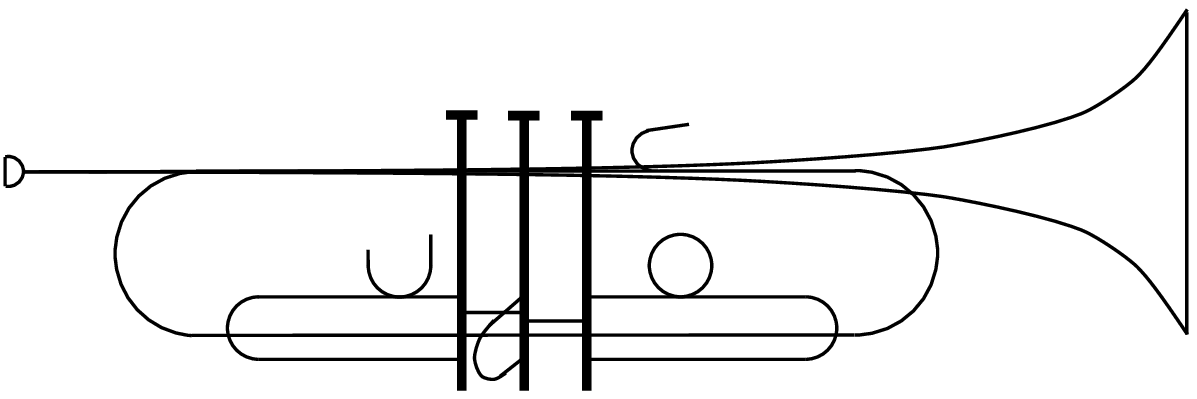,height=0.45in}}}


\bigskip
\bigskip
\begin{center}
   {\Large \bf  Non--Perturbative String Equations for Type 0A}

   \end{center}

\bigskip
\bigskip
\bigskip

\centerline{\bf Clifford V. Johnson\footnote{On leave from the Centre for
    Particle Theory, Department of Mathematical Sciences, University
    of Durham, Durham DH1 3LE, England. \\ ${}^\natural$Address
    for correspondence.}}

\bigskip
\bigskip

  \centerline{\it Department of Physics and Astronomy ${}^\natural$}
\centerline{\it University of
Southern California}
\centerline{\it Los Angeles, CA 90084-0484, U.S.A.}
\centerline{\small \tt johnson1@usc.edu}
\bigskip

\centerline{and}
\bigskip

  \centerline{\it Kavli Institute for Theoretical Physics}
\centerline{\it University of
California}
\centerline{\it Santa Barbara, CA 93106-4030, U.S.A.}
\centerline{\small \tt johnson1@kitp.ucsb.edu}

\bigskip
\bigskip
\bigskip


\begin{abstract}
  Well--defined non--perturbative formulations of the physics of
  string theories, sometimes with D--branes present, were identified
  over a decade ago, from careful study of double scaled matrix
  models.  Following recent work which recasts some of those early
  results in the context of type~0 string theory, a study is made of a
  much larger family of models, which are proposed as type 0A models
  of the entire superconformal minimal series coupled to supergravity.
  This gives many further examples of important physical phenomena,
  including non--perturbative descriptions of geometrical transitions
  between D--branes and fluxes, tachyon condensation, and holography. In
  particular, features of a large family of non--perturbatively stable
  string equations are studied, and results are extracted which
  pertain to type 0A string theory, with D--branes and fluxes, in this
  large class of backgrounds.  For the entire construction to work,
  large parts of the spectrum of the supergravitationally dressed
  superconformal minimal models and that of the gravitationally
  dressed bosonic conformal minimal models must coincide, and it is
  shown how this happens. The example of the super--dressed
  tricritical Ising model is studied in some detail.

\end{abstract}
\newpage \baselineskip=18pt \setcounter{footnote}{0}

\section{Opening Remarks}
\label{sec:introduction}
Between the First and Second Superstring Revolutions, there was
another burst of excitement, based on the technology of large $N$
matrix techniques. This ``Matrix Revolution'' has generally been
regarded as having delivered much less than it seemed to
promise\footnote{Not unlike the recently released film with a similar
  title.}. Occasionally it is still referred to as having taught us a
few of the early lessons about important features of string theory
beyond perturbation theory (see {\it e.g.,}
ref.\cite{Shenker:1990uf}), but further than that not much else has
been said. It has been suggested by some over recent years that there
are more useful lessons to be learned from that period, some of which
may be relevant to current issues (see {\it e.g.,}
ref.\cite{Johnson:1998pc}).  Admittedly, most of the suggestions had
not been bolstered by computational evidence in a definite framework
of ideas, and so were largely ignored. This year has seen a number of
key
papers\cite{McGreevy:2003kb,Martinec:2003ka,Klebanov:2003km,McGreevy:2003ep,Alexandrov:2003nn}
which have supplied both framework and computations renewing interest
in connecting the old results to new ideas, and thereby refurbishing
an old laboratory for the testing of more modern ideas.  In short,
one--dimensional $N{\times}N$ matrix models have been identified as
the world--volume theory of~$N$ of the D--branes the appropriate
continuum theory\cite{Fateev:2000ik} (Liouville coupled to a scalar),
thereby recasting the large~$N$ physics as another tractable example
of holography since the string theory is two--dimensional.
Furthermore, since the matrix model potential is recognised as the
open string world--volume tachyon potential, we have a powerful new
setting within which to study tachyon
condensation\cite{Schomerus:2003vv,Sen:2003iv,Sen:1999mg,Sen:2002nu},
the understanding of which is of considerable importance. For these
and other reasons (see below) it is very clear that the double scaled
technology has delivered on many of its promises and may well continue
to do so as we explore it with fresh eyes\footnote{Sadly, the same
  cannot be said for the film mentioned in the previous footnote.}.

One of the biggest promises on which the double--scaled matrix model
technology is alleged to not have delivered was that of supplying a
complete and well--defined non--perturbative definition of the stringy
physics. This assertion has in fact been known to be untrue, but by
only a few (despite having been present in the literature for a dozen
years).  It is extremely simple and natural\footnote{The
  definition\cite{Dalley:1992qg,Dalley:1992vr,Dalley:1992yi} naturally
  arose first in studies of double--scaled complex matrix models, but
  were quickly shown to be realised in any double scaled matrix model,
  such as unitary or Hermitian, where parameters can be tuned to
  permit the merger of two disconnected components of the Dyson gas of
  eigenvalues. The original complex matrix model scenario, which
  contains a natural ``wall'' into which a single component can be
  tuned to collide, is just a $\IZ_2$ identification of a picture with
  two symmetric components. Matrix models which appeared later (in the
  $c=1$ matrix model context) with such doubled
  distributions\cite{Moore:1992sf,Moore:1992zv,Dhar:1995gw} of
  eigenvalues, have recently been reinterpreted as type~0 string
  theories\cite{Takayanagi:2003sm,Douglas:2003up}.} to define string
theories with exactly the same genus by genus expansion as that
provided by the original matrix
models\cite{Brezin:1990rb,Douglas:1990ve,Gross:1990vs,Gross:1990aw},
but with a perfectly well defined and unique non--perturbative
completion\cite{Dalley:1992qg,Dalley:1992vr,Dalley:1992yi}.

It is also often remarked that it is only recently that we have
achieved understanding of Dirichlet open string sectors ({\it i.e.,}
D--branes) in the context of those early string models. Actually,
perfectly sensible perturbative definitions were derived at an early
stage\cite{Kazakov:1990cq,Kostov:1990nf}, and the non--perturbative
definition of these closed string theories mentioned in the previous
paragraph was extended in ref.\cite{Dalley:1992br} to achieve the
inclusion of world sheets with Dirichlet boundaries. This was used to
show\cite{Dalley:1992br} that the double--scaled unitary matrix
models\cite{Periwal:1990gf,Periwal:1990qb} were not supplying the
physics of some mysterious continuum theory (another misconception
with surprising longevity), but were just a rewriting of string
theories, albeit with the inclusion of open string sectors.

It is worth noting that despite the excitement and remarkable
successes of the Second Revolution, which gave us so much insight into
strings beyond perturbation theory, the bulk of the non--perturbative
information which we have is still phrased very much in terms of
perturbative string theory or field theory, by invoking strong--weak
coupling dualities.  It is still difficult to supply definitions of
physics at intermediate values of the coupling, without restricting
attention to elementary considerations such as the spectra of BPS
states. The $c\leq1$ physics supplied by the double--scaled matrix
model techniques years ago was in terms of families of non--linear
differential equations called ``string equations'' in the old
literature\cite{Brezin:1990rb,Douglas:1990ve,Gross:1990vs,Gross:1990aw}.
The solutions of these equations obtained by asymptotic expansion
correspond to weak string coupling perturbation theory, but they
encode physics at arbitrary values of the coupling in the full
solutions. These systems may well contain useful lessons about string
theory at intermediate coupling which may help address issues of
current concern.

The reason for recalling the various facts listed above is 
so that we may exhibit them as components of a possible series of
further valuable lessons which the Matrix Revolution taught us which
were missed the first time around. Perhaps it was necessary for the
field to depart and return to the results from another direction,
enriched with the lessons learned elsewhere, such as the importance of
branes in getting access to strongly coupled physics, the significance
of large $N$ in other stringy contexts such as holography, {\it etc}.
Quite a few of these features are evident in the double scaled matrix
models, but of course they are hard to see because there is not much
room for physics to happen in these low dimensional scenarios.

A recent paper\cite{Klebanov:2003wg} has gathered considerable
evidence for a picture in which many of the aforementioned facts fit
together in a way which is highly instructive, and quite exciting.
First, the string theories for which the well--defined
non--perturbative formulation was provided are not the simple bosonic
string theories but in fact the type~0 string theories (although there
are key similarities; see later).  Specifically, the type 0B string
theories of the $(2,2m)$ superconformal minimal models coupled to
super--Liouville theory are captured by unitary one--matrix models (or
equivalently two--cut Hermitian one--matrix models), while the type 0A
theories of the same worldsheet models are captured by the complex
one--matrix models (or symmetric two--cut Hermitian one--matrix
models, or equivalently the sector of even perturbations of the
unitary one--matrix models). The physics of these models is captured
by string equations which supply unique and sensible non--perturbative
data. In this new context, some of the new lessons we might learn from
matrix models immediately begin to become apparent. For example, the
interpretation of the inclusion of the open string sectors already
supplied in the earlier work\cite{Dalley:1992br} becomes extended to
include Ramond--Ramond flux as well, and the non--perturbative physics
supplied by the string equations connects the two, since they arise as
different asymptotic expansions of the same equation. In fact, such an
elegant non--perturbative encapsulation of geometric
transitions\cite{Gopakumar:1998ki,Klebanov:2000hb,Maldacena:2000yy,Vafa:2000wi,Cachazo:2001jy}
between open string (D--branes) and closed string (fluxes) physics is
highly sought after in the modern context, and for at least that
reason it is worth revisiting the old models, now dressed in their new
clothes.

\subsection{The Purpose of this Paper}

The purpose of the work reported on in this paper is to begin to
define and explore more models in this new context, and study new and
more complicated examples. As stated above, the work of
refs.\cite{Dalley:1992qg,Dalley:1992vr,Dalley:1992yi} in defining
non--perturbative string physics for all the $(2,2k-1)$ conformal
models coupled to Liouville theory, extended further in
ref.\cite{Dalley:1992br} to naturally include open string sectors in
the non--perturbative discussion while also interpreting doubled
unitary scaled matrix models in these terms, was recently recast in
terms of type 0A string theories (with maps to type~0B in (at least)
one case, using the result of ref.\cite{Morris:1990bw,Morris:1992zr}),
where the relevant conformal models are the $(2,2m)$ superconformal
series, where $m=2k$.  In particular, for type~0A, the string equation
is that of
refs.\cite{Morris:1992zr,Dalley:1992qg,Dalley:1992vr,Dalley:1992yi}:
\begin{equation}
u{\cal R}^2+\frac{1}{2}{\cal R}
{\cal R}^{\prime\prime}-\frac{1}{4}({\cal R}^\prime)^2=\nu^2\Gamma^2\ 
,\qquad {\rm where}\,\,\, f^\prime\equiv \nu{\partial f\over \partial
  z},
  \label{nonperturbative}
\end{equation}and $\nu$ is a parameter related to the closed string
coupling, since it is the renormalised $1/N$ which survives the double
scaling limit of the ($N{\times}N$) matrix model. Furthermore,
\begin{equation}
{\cal R}=\sum_{l=1}^{k}\left(1+\frac{l}{2}\right)t_l R_l[u] -z\ ,
  \label{Rdef}
\end{equation}
where the $t_l$, $l=1,\ldots,k$, are couplings to the gravitationally
dressed conformal field theory operators, ${\cal O}_l$, in the theory
and the $R_l[u]$ are polynomials in $u$ and its derivatives with
respect to~$z$, given below in equation~\reef{eq:gelfanddikki}. The
function $u$ is the two point function of the operator ${\cal O}_0$,
which couples to $z=-t_0$, and so the free energy  can be
obtained by integrating $u$, as the precise relation is:
\begin{equation}
u=-2\nu^2\frac{\partial^2 F}{\partial z^2}\ .
  \label{partfun}
\end{equation}
In ref.\cite{Dalley:1992br}, it was shown that the large positive~$z$
expansion of a solution to equation~\reef{nonperturbative} has an
interpretation in terms of string world--sheet perturbation theory in
the presence of $\Gamma$ branes, since $\Gamma$ appears raised to the
power $b$, where $b$ is the number of boundaries in the genus
expansion. For $\Gamma=0$, the perturbation expansion is identical to
that of the string equations of
refs.\cite{Brezin:1990rb,Douglas:1990ve,Gross:1990vs,Gross:1990aw}
which have the ill--defined non--perturbative physics for $k$ even.
Indeed, those equations are simply contained in the above equation as
the case ${\cal R}=0$. For example, in the case of pure gravity, the
$k=2$ model, we have:
\begin{equation}
u(z)=z^{\frac12} +\frac{1}{2}\frac{\nu\Gamma}{z^{{\frac{3}{4}}}}
-\frac{1}{24}\frac{\nu^2}{z^2}(6\Gamma^2+1)
+\frac{15}{256}\frac{\nu^3}{z^{\frac{13}{4}}}\Gamma(4\Gamma^2+3)-\frac{7}{4608}\frac{\nu^4}{z^{\frac92}}(180\Gamma^4+345\Gamma^2+28)+\cdots
  \label{eq:expand}
\end{equation}
On integrating to get the free energy, it is seen that the natural
dimensionless topological expansion parameter (closed string coupling)
is $\kappa=\nu/z^{5/4}$. Worldsheets with $h$ handles and~$b$
boundaries come with a factor $\kappa^{2h-2+b}\Gamma^b$, and so
$\Gamma$ naturally has an interpretation as the number (in some
units\footnote{In fact, for the purposes of this paper, the precise
  normalisation of $\Gamma$ does not matter too much. It is enough to
  know that $\Gamma$ simply counts the number of D--branes, or half
  R--R flux insertions (see later), in some units. All that is
  important here is its exponent in various equations, which help
  organise perturbation theory.  This was why $\Gamma$ was called the
  ``open string coupling'' in ref.\cite{Dalley:1992br}.})  of
D--branes to which each boundary couples\footnote{The observation in
  ref.\cite{Dalley:1992br} that double scaled unitary models of
  ref.\cite{Periwal:1990gf,Periwal:1990qb} are just the usual string
  theories with open string sectors was based on realising that
  solutions of the string equations obtained in that context can be
  mapped to solutions of equation~\reef{nonperturbative}, with the
  boundary conditions such that their expansion is of the
  form~\reef{eq:expand}, for a specific value of $\Gamma$.}.

That the equation~\reef{nonperturbative}, for $\Gamma=0$, has a
sensible large negative~$z$ expansion, in contrast to the case ${\cal
  R}=0$, was key to the realisation of
refs.\cite{Dalley:1992qg,Dalley:1992vr,Dalley:1992yi} that there is
sensible non--perturbative formulation.  In fact, for arbitrary
$\Gamma$, the $z\to-\infty$ expansion has a sensible world--sheet
interpretation, and the regions of intermediate~$z$ connect smoothly
through $z=0$ to each other in a unique\footnote{Uniqueness is
  suggested by numerical
  work\cite{Dalley:1992qg,Dalley:1992vr,Dalley:1992yi}, and by direct
  proof in special cases\cite{Johnson:1992pu}.  Furthermore, setting
  $\Gamma=1/2$, this equation can be mapped using the Miura map to the
  family of solutions to the string equations of the double--scaled
  unitary matrix models with even potentials. These are known to be
  unique from analytic studies. See also the previous footnote.} way,
supplying a complete non--perturbative definition of the physics. For
example, for $k=2$ again, we have:
\bigskip
\bigskip

\begin{eqnarray}
u(z)&=&\frac{1}{4}\frac{\nu^2}{z^2}(4\Gamma^2-1)
+\frac{1}{32}\frac{\nu^6}{z^7}(4\Gamma^2-1)(4\Gamma^2-9)(4\Gamma^2-25)\times\nonumber\\&&\hskip5cm\left(1+\frac{11}{96}\frac{\nu^4}{z^{5}}(48\Gamma^4-1240\Gamma^2+8371)\right)+\cdots
  \label{eq:expandmore}
\end{eqnarray}
In the interpretation in terms of open strings, it is strange that
there are no world sheets with an odd number of boundaries, and no
interpretation was given in the earlier work for the missing orders,
although a connection with the ``topological'' solution
$u=\nu^2/(4z^2)$ was suggested, since for all $k$ this is the leading
term in all the large negative $z$ expansions. In the new setting of
ref.\cite{Klebanov:2003wg}, this large negative~$z$ expansion has a
new interpretation: The string theory is closed and $\Gamma$
represents an even number of insertions due to the presence of
$\Gamma$ units of R--R flux. Since the two expansions ---one with
D--branes and one with R--R flux--- are connected smoothly through the
strong coupling region by the full string
equation~\reef{nonperturbative}, this gives a complete and elegant
example of geometric transitions between descriptions with D--branes
and those with R--R fluxes.

There is a large family of natural generalisations of the string
equation \reef{nonperturbative}. These were presented in
refs.\cite{Johnson:1992uy}, where the context was the full family of
$(q,p)$ conformal minimal models coupled to Liouville theory. The
principal techniques used there were the underlying structure of the
generalised KdV hierarchy which organises the physics. The inclusion
of open string sectors was derived in ref.\cite{Johnson:1994vk}, by
exploiting further the underlying integrable system, and appealing to
the elegant organisation of the key structures in terms of integrable
systems and the modes of a twisted boson. Lest fears arise that a loss
of contact with a matrix model and hence a worldsheet description may
have resulted from paying too much attention to the integrable system
to derive these generalisations, it should be noted that the
open--closed string equation of ref.\cite{Johnson:1994vk} for the case
$(4,3)$ was reproduced in an explicit two--matrix model computation in
ref.\cite{Houart:1993nq}.

In this paper these more general equations, presented in
section~\ref{proposal}, will be explored somewhat further than they
were when originally presented, in the light of the new setting. We
will make a number of observations about the physics they contain.
Part of the goal is to make the natural suggestion that they are the
equations defining the type~0A system of the full $(q,p)$
superconformal minimal models coupled to super--Liouville theory. For
this to work, it is evident that there must be a striking coincidence
between the spectrum of dressed operators in the superconformal theory
and those in the bosonic theory, for which the string equations were
originally presented. This coincidence was already made explicit for
the $(2,2k-1)$ (bosonic) {\it vs.} the $(2,4k)$ (supersymmetric) case
presented in ref.\cite{Klebanov:2003wg}, and it is natural to
conjecture that it is more generally true. In
section~\ref{superconformal} we show how it works in general.

In section~\ref{examples}, we study a specific example, which is in
fact to be interpreted as the tricritical Ising model not with the
ordinary gravitational dressing, but {\it super}gravitationally
dressed\footnote{In fact, it now wears a cape.}.  We observe some
features of the perturbative expansions and anticipate that the full
non--perturbative solution is well defined. Our observations strongly
suggest that there is a rich family of geometrical transitions between
D--branes and R--R fluxes, all of which the model has a complete
non--perturbative description. There also seem to be possible
solutions with a novel interpretation in terms of R--R flux in {\it
  both} weak coupling regimes for which D--branes can be switched on
in one weak coupling region by switching on a further operator. This
operator induces a flow to an open string generalisation of what is
the ordinary Ising model in the bosonic context. We note that more
work is needed to study the non--perturbative solutions of the string
equations, since our computations have been centered around exploring
the possible weak coupling limits of the string equation, and
anticipating that at least a subset of the behaviours that we find are
connected to each other by the non--perturbative physics contained in
the equation.  Such further study to explore the fate of the solutions
we have found is likely to be highly instructive and fruitful.

Finally, the non--trivial numbers appearing in the one--loop partition
function of the $q=3$ series of models are noted, since they are clues
to confirming the identity of the models. Unfortunately, we cannot
conclude anything about them until the corresponding numbers for the
type 0B model are known, since the contributions from the two sets of
spin structures must be added to compare with known results.
Formulating the $(q,p)$ 0B models is beyond the scope of this paper,
and so a full comparison is left to the future.  There is an
interesting feature however, extending what was observed for the $q=2$
case long ago\cite{Dalley:1992br}, and interpreted in the present
context in ref.\cite{Klebanov:2003wg}: A large class of solutions with
flux interpretation in the large $z$ expansion all have vanishing
contribution on the sphere, and the value of the torus contribution is
universal to all models in the 3--series. This is likely to be true in
the general $(q,p)$ string models.

\section{More General String Equations}
\label{proposal}
The purely closed string $(q,p)$ generalisations of the $(2,2k-1)$
string equation~\reef{nonperturbative} were presented in
ref.\cite{Johnson:1992uy}, although the fullest discussion of
properties and structure of the entire family of equations and their
associated matrix models can be found in ref.\cite{Johnson:thesis}.  A
summary follows, and then the open string generalisation, derived in
ref.\cite{Johnson:1994vk} will be stated.

\subsection{Closed Strings}
To appreciate the structure of the equations, it is most useful to 
have in mind the structure of the generalised KdV hierarchies. The
reader not interested in this structure can simply skip to the bottom
of this subsection. 

The $q$th KdV hierarchy can be formulated in terms of  a $q$th order
differential operator $Q$, of the form:
\begin{equation}
Q=d^q+\sum_{i=2}^q \alpha_i\{u_i,d^{q-i}\}\ ,
  \label{eq:Qdiffy}
\end{equation}
where there are $q-2$ functions $u_i$, and $d$ here denotes ($\nu$
times) the derivative with respect to~$z$. The $\alpha_i$s are
normalisation constants.  The hierarchy is a family of flow equations:
\begin{equation}
\frac{\partial Q}{\partial t_r}=[Q^{\frac{r}{q}}_+,Q]\ ,
  \label{eq:Qflows}
\end{equation}
where $Q^{1/q}_+$ is constructed using the elegant technology of
pseudo--differential operators\cite{Gelfand:1976A}. Briefly,
$Q^{\frac{1}{q}}$ is an infinite series:
\begin{equation}
Q^{\frac1q}=d+\sum_{j=1}^{\infty}S_j[u_i] d^{-1-j}\ ,
  \label{eq:Qfrac}
\end{equation}
where the operation $d^{-1}$ is defined by:
\begin{equation}
d^{-1}u=\sum_{i=0}^\infty (-1)^i u^{(i)}d^{-1-i}\ ,
  \label{eq:dminusone}
\end{equation}
where $u^{(i)}$ means $u(z)$ differentiated $i$ times with respect to
$z$. The functions $S_j$ in the above are polynomials in the $u_i$
and their~$z$--derivatives.  With the above definition, is easy to
raise $Q^{\frac1q}$ to the desired integer power $r$, and then
$Q^{\frac{r}{q}}_+$ denotes the truncation of the result by discarding
terms containing odd powers of $d$.

Of course, if $r=0$ mod $q$, the flows defined in
equation~\reef{eq:Qdiffy} are trivial, and so we only consider the
case when $q$ does not divide $r$. Hence, we write $r=qk+l$, where
$l=1,\ldots,q-1$, and $k=0,\ldots,\infty$. The generalised times $t_r$
will be written $t_{l,k}$.

Just for orientation, the more familiar KdV hierarchy is the case
$q=2$, for which there is just one index to vary, $k$, which gives the
$k$th flow for the function $u_2=-u$:
\begin{equation}
\frac{\partial u}{\partial t_k}={\cal R}_{k+1}^\prime\ ,
  \label{eq:kdvh}
\end{equation}
where the ${\cal R}_k$ are given in equation~\reef{eq:gelfanddikki}.
So in the more general case of $q>2$, one can think of~$l$ as
parameterising $q-1$ different towers of flows, where $k$ denotes
where one is on the tower.

After some work, one can write the general flow equations directly as
differential equations for the $u_i$:
\begin{equation}
\alpha_{i}\frac{\partial u_i}{\partial
  t_{l,k}}={\cal D}_1^{ij}R^j_{l,k+1}={\cal D}_2^{ij}R^j_{l,k}\
  ,\qquad i,j=2,3,\ldots q\ ,
  \label{eq:Qdiffytwo}
\end{equation} where there's no sum on $i$ intended, and the
$\alpha_i$ are constants. The derivative operators ${\cal D}_{1,2}$
arise from the ``bi--Hamiltonian'' Poisson bracket structure of the
the KdV systems, and define recursion relations between the
differential polynomials $R^i_{l,k}$ {\it via} the second relation in
equation~\reef{eq:Qdiffytwo}. These polynomials in the $\{u_i\}$ and
their derivatives are generalisations of the famous
Gel'fand--Dikii\cite{Gelfand:1976B} polynomials which appear in the
usual KdV system, where $Q=d^2-u$, and:
\begin{eqnarray}
&&
R_0=2\ ,\quad R_1=-u\ ,\quad
R_2=\frac14\left(3u^2-u^{\prime\prime}\right)\ , \nonumber\\
&&
R_3=-\frac{1}{16}\left(10u^3-10uu^{\prime\prime}-5(u^\prime)^2+u^{(4)}\right)\
,\cdots\ ,
  \label{eq:gelfanddikki}
\end{eqnarray}
and the derivative operators in this $q=2$ case are:
\begin{equation}
{\cal D}_1=d\ ,\qquad {\cal D}_2=\frac{1}{4}d^3-\frac{1}{2}u^\prime-u
d\ ,
  \label{eq:derivatives}
\end{equation}
so that ${\cal D}_1R_{k+1}={\cal D}_2R_k$. In general, the $R^i_{l,k}$
are fully determined by acting with the ${\cal D}_{1,2}$ operators,
once the seed constants are fixed. In our normalisation,
$R^j_{1,0}=q\delta_{l}^{j-1}$. An explicit example of the case $q=3$
showing how the more general structure works, will be given shortly.

With all of the structures in place, it is quite simple to state the
general form of the string equations. The following set of $q-1$  equations:
\begin{equation}
{\cal D}_2^{ij}{\cal R}^j=0\ ,
  \label{eq:generalstring}
\end{equation}
where 
\begin{equation}
{\cal  R}^i=
\sum_{l=1}^{q-1}\sum_{k=0}^{\infty}\left(k+\frac{l}{q}\right)t_{l,k}
R^i_{l,k}=
\sum_{l=1}^{q-1}\sum_{k=1}^{\infty}\left(k+\frac{l}{q}\right)t_{l,k}
R^i_{l,k}+(i-1)t_{i-1,0}\ ,
  \label{eq:defineR}
\end{equation}
may be combined into a single equation by multiplying on the left by
${\cal R}^i$ and summing. The result is in fact a total derivative.
The equation resulting from integrating once and setting the constant
to zero is the string equation\footnote{In fact, the string equations
  of
  ref.\cite{Brezin:1990rb,Douglas:1990ve,Gross:1990vs,Gross:1990aw},
  for which the general $(q,p)$ case was formulated in terms of
  integrable hierarchies in ref.\cite{Douglas:1990dd}, can be written
  as the first integrals of the $q-1$ equations ${\cal D}_1^{ij}{\cal
    R}^j=0$. The $t_{i-1,0}$ arise as integration constants, and so in
  our conventions those string equations are just ${\cal R}^j=0$. The
  non--perturbative problems of these equations are well--known. In
  the present formulation, these equations are taken as only
  perturbatively true, the correct non--perturbative physics being
  supplied by equation~\reef{eq:stringequationfinal}.}:
\begin{equation}
\int\!\! dz \,{\cal R}^i {\cal D}^{ij}_2{\cal R}^j=0\ .
  \label{eq:stringequationfinal}
\end{equation}

This elegant compact form may not be totally illuminating to the
reader, so it is worthwhile to see how the $q=2$ string
equation~\reef{nonperturbative} arises. It can be done by eye, looking
at ${\cal D}_2$ defined in~\reef{eq:derivatives}, with ${\cal R}$
given in~\reef{Rdef}, one gets a total derivative, ${\cal R}{\cal
  D}_2{\cal R}$, which can be integrated once to give
equation~\reef{nonperturbative}. To match onto closed string
perturbation theory, which is equivalent to ${\cal R}=0$, the constant
must be set to zero.

\subsection{Open Strings}

As might be guessed from an examination of the $q=2$ equations already
stated, the generalisation to include some number, $\Gamma$, of
D--branes is straightforward. One must simply keep non--zero the
integration constant at the last step of the previous subsection! So
the general string equation, which includes world sheet boundaries in
the $z\to+\infty$ perturbative expansion, is:
\begin{equation}
\int\!\! dz \,{\cal R}^i {\cal D}^{ij}_2{\cal R}^j=\nu^2\Gamma^2\ .
  \label{eq:stringequationfinalfinal}
\end{equation}

The reader might find the above open string formulation a little
brief, and thus find it hard to believe that it can be so simple to
include D--brane sectors. There is in fact an elegant underlying
structure to all of this which is being suppressed.  First, from a
direct matrix model computation, for the $(2,2k-1)$ case the
perturbative open string physics can be explicitly recovered by
including a term $\frac{\Gamma}{N}\log (1-M^2)$ in the one--matrix
model of the $N\times N$ matrix $M$, as was carried out originally in
ref.\cite{Kazakov:1990cq,Kostov:1990nf}. The result is in fact:
\begin{equation}
{\cal R}=\frac{1}{2}\nu\Gamma {\hat R}\ , 
  \label{eq:openperturb}
\end{equation}
where ${\hat R}$ is the resolvent of the differential operator
$Q=d^2-u$, defined as a formal inverse\cite{Gelfand:1975rn,Gelfand:1976B}:
\begin{equation}
{\hat R}(z,\sigma)\cdot (Q-\sigma)=1\ ,\quad{\it i.e.,}\quad {\hat
  R}(z,\sigma)=<\!z|\frac{1}{\nu^2\partial^2_z-u-\sigma}|z\!>\ .
  \label{eq:resolvent}
\end{equation}
In fact, the differential polynomials are simply coefficients in the
expansion in the spectral parameter $\sigma$:
\begin{equation}
{\hat R}(z,\sigma)=\sum_{k=0}^\infty
\frac{R_k[u]}{\sigma^{k+\frac{1}{2}}}\ ,
  \label{eq:expandR}
\end{equation}
and the resolvent itself satisfies a differential equation\cite{Gelfand:1975rn}:\begin{equation}
(u+\sigma){\hat R}^2+\frac{1}{2}{\hat R}
{\hat R}^{\prime\prime}-\frac{1}{4}({\hat R}^\prime)^2=4\ ,
  \label{eq:resolventdiffy}
\end{equation}
which is in fact the first integral of the equation ${\hat R}({\cal
  D}_2+\sigma{\cal D}_1){\hat R}=0$, and the integration constant is
set by the normalisation of the $R_k$.  Henceforth, we will set the
spectral parameter $\sigma$ to zero\footnote{In fact, $\sigma$ has
  considerable significance in this context, as it couples to a
  boundary operator.  Its effects are
  interesting\cite{Johnson:1992wr}, but we shall not explore them
  here.}. So one can recover the full perturbation theory of solutions
of equation~\reef{eq:openperturb} by simply expanding the resolvent to
the desired order in equation~\reef{eq:resolventdiffy}.  The
non--perturbative physics comes from the full equation, obtained
simply by solving equation~\reef{eq:openperturb} for the resolvent and
substituting into equation~\reef{eq:resolventdiffy}. The result is of
course our equation~\reef{nonperturbative}, and can in fact be derived
directly in an appropriately tuned complex matrix model or two--cut
Hermitian matrix model.

The general structure was recognised in ref.\cite{Johnson:1994vk} and
exploited to derive the full open string generalisation above. Briefly
explained, the relevant resolvent defines a family of ``jet
coefficients'', ${\hat R}^i$ which are like $q-1$ generalisations of
the single resolvent ${\hat R}$ from before\cite{Gelfand:1976B}.
There is a resolvent equation which is simply the first integral of
${\hat R}^i\left({\cal D}_2^{ij}+\sigma{\cal D}_1^{ij}\right){\cal
  R}^j=0$, where the integration constant is in fact~$q^2$ in the
normalisation we chose earlier. The open string generalisation of
perturbation theory is simply ${\cal R}^j=\nu\Gamma {\hat R}^j/q$, and
the full non--perturbative physics follows again by using the larger
equation after substituting for the resolvent, which is what we wrote
in equation~\reef{eq:stringequationfinalfinal}. Much more about the
structure of the equations can be found in ref.\cite{Johnson:1994vk}.
Note also that an explicit two--matrix model realisation of an example
of this more general formulation was derived explicitly in
ref.\cite{Houart:1993nq}.

\subsection{The Proposal}
The proposal is simple. The string equations presented above are to be
taken as a definition of the type 0A string theory in less than two
dimensions, in a particular family of backgrounds.  In particular the
backgrounds are the $(q,p)$ superconformal minimal models, and the
``non--critical'' stringy physics is realised by coupling them to
super--Liouville theory.  The Liouville coupling is done through the
lowest dimension operator in the theory, with coupling given by~$z$.
In the large positive~$z$ perturbative world--sheet expansion, there
will be boundaries, and $\Gamma$ represents the number of background
D--branes in the model. It is expected that large negative~$z$ will
give a closed string expansion with insertions of $\Gamma$ units of
R--R flux.  The string equations have smooth (and probably unique in
general) solutions which connect these asymptotic behaviours to each
other through the non--perturbative region of small~$z$.

The bulk operators in the theory are readily identified in the $q$--KdV
formalism. One picks a particular theory by picking $L$ and $K$ such
that $Kq+L=p$ and for this choice, set
\begin{equation}
\left(K+\frac{L}{q}\right)t_{l,k}=\delta_l^L\delta_k^K.
  \label{eq:setting}
\end{equation}
All higher $t_{l,k}s$ are set to zero. All the lower ones will be
couplings to operators\footnote{One of them is in fact a boundary
  operator and not a bulk field at all\cite{Martinec:1991ht}.}. In
particular, the family of $q-1$ operators $t_{l,0}$ are very natural
in the theory, since $z=-t_{1,0}$ is the coupling to the lowest
dimension operator, and its cousins ${\cal B}_i=-t_{i,0}$ have an
interpretation as a family of generalised fields such as the magnetic
field (see below).  We shall study some examples below.

The evidence for the case $q=2, p=2m$ was already presented in
ref.\cite{Klebanov:2003wg}.  This proposal extends that work further
to all of the superconformal minimal models in a natural way, and it
will certainly be interesting to study the role (in this new context)
of the rich structures which determine the theory's internal
organisation, such as the generalised $q$--KdV flows, the
$W^{(q)}$--algebra of constraints, the language of twisted bosons,
{\it etc}.

\section{Dressed Superconformal Minimal Models and their Bosonic Cousins}
\label{superconformal}
In proposing that the non--perturbative string physics found in the
string equation~\reef{nonperturbative} actually pertains to the
type~0A string theory, the operators of the bosonic string theory (the
$(2,2k-1)$ conformal minimal models coupled to gravity) had to be
mapped onto the operator content of the $(2,4k)$ superconformal
minimal models, dressed with gravity. This was carried out in
ref.\cite{Klebanov:2003wg}. This is in itself an interesting fact, and
if the proposal of this paper is to be believed, we must show that
this is the case more generally since there is the danger that the
correspondence stops with that sub--series.

First, a brief reminder of the bosonic case is in order. (A review is
given in ref.\cite{Ginsparg:1993is}, for example.) Consider the
$(q,p)$ theory. It has central charge given by
\begin{equation}
c=1-6\frac{(p-q)^2}{pq}\ .
  \label{eq:centralcharge}
\end{equation}
The bare operators are labelled by two integers, $r$ and $s$,
satisfying $1\leq r\leq q-1$, $1\leq s \leq p-1$, with $pr\geq qs$, and
shall be denoted ${\cal O}^0_{r,s}$. Their dimensions in ordinary
conformal field theory (``bare dimensions'') are:
\begin{equation}
h_{r,s}=\frac{(pr-qs)^2-(p-q)^2}{4pq}\ .
  \label{eq:conformaldimension}
\end{equation}
After coupling to gravity, the operator is dressed by the Liouville
field $\varphi$ according to 
\begin{equation}
{\cal O}_{r,s}=e^{\beta_{r,s}\varphi}{\cal
  O}_{r,s}^0\ ,
  \label{eq:dressup}
\end{equation}
 and conformal invariance determines the dressing to be:
\begin{equation}
\beta_{r,s}=\frac{p+q-(pr-qs)}{\sqrt{2pq}}\ .
  \label{eq:dressing}
\end{equation}
It turns out that the physics of the operators which arise from the
doubled scaled approach is as follows: In unitary models ($p=q+1$),
the Liouville coupling is through the cosmological constant, which
couples to the area while
more generally, it is to the smallest dimension operator in the
theory. From the above, this is when $pr-qs=1$, and so the Liouville
dressing of this operator is $\beta=(p+q-1)/\sqrt{2pq}$. The dimension
which we really care about for the operator ${\cal O}_{r,s}$ is its
Liouville scaling relative to the smallest dimension operator. This
is:
\begin{equation}
\Delta_{r,s}=\frac{\beta_{r,s}}{\beta}=\frac{p+q-(pr-qs)}{p+q-1}\ .
  \label{eq:liouvilledimension}
\end{equation}

Turning to the superconformal $(q,p)$ series, the operators are
labelled in the same way as before. However, there are two sectors: If
$r-s$ is even the operator is from the NS sector while it is from the
R sector otherwise. The central charge is:
\begin{equation}
{\hat c}=1-2\frac{(p-q)^2}{pq}\ ,
  \label{eq:centralagain}
\end{equation}
and the bare weights of operators are:
\begin{equation}
h_{r,s}=\frac{(pr-qs)^2-(p-q)^2}{8pq}+\frac{1-(-1)^{p-q}}{32}\ .
  \label{eq:nakedwieghts}
\end{equation}
The Liouville weights are 
\begin{equation}
\beta_{r,s}=\frac{p+q-(pr-qs)}{2\sqrt{pq}}\ .
  \label{eq:superdressing}
\end{equation}
Note that the R sector operator with vanishing $pr-qs$ is in fact the
R ground state, with the familiar weight $h={\hat c}/{16}$.  In
identifying the lowest dimension operator for the next step, it is
important to distinguish between the two sectors, and to note the
restrictions on the allowed $(q,p)$. The point is that $p>q$, and $p$
and $q$ are either both odd or both even. In the odd case, they must
be coprime, while in the even case $p/2$ and $q/2$ must be coprime. So
the lowest dimension operator in the case when they are even comes
from the NS sector, with $\beta=(p+q-2)/\sqrt{2pq}$. This was the case
identified in ref.\cite{Klebanov:2003wg}, whence they further
specialised to the case $(q=2,p=4k)$, and recovered the same spectrum
as the $(2,2k-1)$ bosonic minimal models.  The general formula for
even $(q,p)$ for the relative Liouville weight of operator ${\cal
  O}_{r,s}$ is:
\begin{equation}
\Delta_{r,s}=\frac{p+q-(pr-qs)}{p+q-2}\ ,
  \label{eq:superliouvilledimension1}
\end{equation}
from which, after cancelling various factors of two, seeing that $r=1$
only, gives the familiar result  for the scaling of the operator
${\cal O}_{j}$ coupling to $t_{j}$ in the  $k$th model:
\begin{equation}
\Delta_j=1-\frac{j}{k}\ .
  \label{eq:familiardressing}
\end{equation}
The formula of equation~\reef{eq:superliouvilledimension1} stands for
all the even cases, and it should be clear that there will always be
factors of two cancelling on the top and bottom lines to allow an
embedding of the bosonic operator
spectrum~\reef{eq:liouvilledimension} into the supersymmetric one.
There remains the case of odd $(q,p)$.  Here the lowest dimension
operator is in fact from the R sector, and we see that
$\beta=(p+q-1)/\sqrt{2pq}$.  Therefore the relative Liouville weight
of operator ${\cal O}_{r,s}$ is:
\begin{equation}
\Delta_{r,s}=\frac{p+q-(pr-qs)}{p+q-1}\ ,
  \label{eq:superliouvilledimension2}
\end{equation}
There are no cancellations of factors of 2 required here, and we see
that this has exactly the same form as the generic bosonic
formula~\reef{eq:liouvilledimension}. This is an interesting and
fortuitous fact which gives weight to the main proposal of this paper.

In what follows, we shall focus on a particular set of examples to
check our understanding, and the precise content of the proposal.

\section{The 3--Series}
\label{examples}
We are now ready to unpack a particular example and play with it. Let
us focus on the case $(3,p)$. The objects we need are the differential
polynomials from the 3--KdV (or Bousinnesq) hierarchy, or
equivalently, the structures ${\cal D}^{ij}_{1,2}$ which can be used
to generate them by recursion starting from $R^2_{1,0}=3=R^3_{2,0},
R^2_{2,0}=0=R^3_{1,0}$.

Choosing
\begin{equation}
Q=d^3+\frac{3}{4}\left\{u_2,d\right\}+u_3\ ,
  \label{eq:Qchoice}
\end{equation}
we have:
\begin{eqnarray}
{\cal D}^{22}_2&=&\frac{2}{3}d^3+\frac{1}{2}u_2^\prime+u_2 d\ , \qquad
  {\cal
  D}_2^{23}=u_3 d +\frac{2}{3}u_3^\prime\ , \qquad {\cal D}^{32}_2= u_3 d
  +\frac{1}{3}u_3^\prime\ , \nonumber\\
{\cal D}^{33}_2&=&-\frac{1}{18}d^5-\frac{5}{12}u_2 d^3
  -\frac{5}{8}u_2^\prime d^2+\left(-\frac{1}{2}u_2^2-\frac{3}{8}u_2^{\prime\prime}\right) d +\left(-\frac{1}{2}u_2u_2^\prime-\frac{1}{12}u_2^{\prime\prime\prime}\right)\ ,
  \label{eq:differentialstructures1}
\end{eqnarray}
and, more simply
\begin{eqnarray}
{\cal D}_1^{22}=0={\cal D}_1^{33}\ ,\qquad {\cal D}_1^{23}=d={\cal
  D}_1^{32}\ .
  \label{eq:differentialstructures2}
\end{eqnarray}

We will need the explicit expressions for the first three rungs
$k=0,1,2$ on the two ladders making up the hierarchy of $R^i_{l,k}$ in
order to construct our examples. They can be determined to be (after
two applications of the ${\cal D}_2$ operator according to the right
hand side of equation~\reef{eq:Qdiffytwo}, integrating the resulting
total derivatives each time):
\begin{eqnarray}
&&R^2_{1,0}=3\ ,\qquad R^3_{1,0}= 0 \ , \nonumber\\
&&R^2_{2,0}=0\ ,\qquad R^3_{2,0}=3 \ ;\nonumber\\
&&R^2_{1,1}=u_3\ ,\qquad R^3_{1,1}=\frac{3}{2}u_2\ ,\nonumber\\
&&R^2_{2,1}=-\frac{1}{4}\left(u_2^{\prime\prime}+3u_2^2\right)\ ,\qquad
R^3_{2,1}=2u_3\ ;\nonumber\\
&&R^2_{1,2}=-\frac{1}{12}u_2^{''''}-\frac{3}{4}u_2u_2^{''}-\frac{3}{8}(u_2^\prime)^2-\frac{1}{2}u_2^3+\frac{2}{3}u_3^2\
,\quad R^3_{1,2} =\frac{2}{3}\left(u_3^{\prime\prime}+3u_3u_2\right)\
,\nonumber\\
&&R^2_{2,2}=-\frac{1}{9}u_3^{''''}-\frac{5}{6}u_2u_3^{\prime\prime}-\frac{5}{12}u_2^{\prime\prime}u_3-\frac{5}{12}u_2^\prime
u_3^\prime-\frac{5}{4}u_2^2u_3\ ,\nonumber\\
&&\hskip5cm
R^3_{2,2}=-\frac{1}{6}u_2^{''''}-\frac{5}{4}u_2u_2^{\prime\prime}-\frac{15}{16}(u_2^\prime)^2 -\frac{5}{8}u_2^3+\frac{5}{3}u_3^2\ .
  \label{eq:differentialpolynomials}
\end{eqnarray}

The string equation is rather long when written out, and
is\cite{Johnson:1994vk}:
\begin{eqnarray}
&&\frac{1}{2}u_2{\cal R}_2^2+\frac{2}{3}{\cal R}_2{\cal
  R}_2^{\prime\prime}-\frac{1}{3}\left({\cal
  R}_2^\prime\right)^2+u_3{\cal R}_2{\cal R}_3-\frac{1}{18}\left({\cal
  R}_3{\cal R}_3^{''''}-{\cal R}_3^\prime{\cal
  R}_3^{'''}+\frac{1}{2}\left({\cal
  R}_3^{\prime\prime}\right)^2\right)\nonumber\\
&&-\frac{5}{12}\left(u_2{\cal R}_3{\cal
  R}_3^{\prime\prime}-\frac{1}{2}u_2\left({\cal
  R}_3^\prime\right)^2+\frac{1}{2}u_2^\prime{\cal R}_3{\cal
  R}^\prime_3\right)
-\frac{1}{12}\left(3u_2^2+u_2^{\prime\prime}\right){\cal R}_3^2=\nu^2\Gamma^2\ .
  \label{eq:thestringequation}
\end{eqnarray}
The indices on the ${\cal R}^i$ have been temporarily dropped from
upper to lower to permit a tidier equation to be written. The content
of the ${\cal R}^i$ will be determined by which model is chosen.

\subsection{The (3,4) (Bosonically Dressed Ising) model as a subsector.}
Actually, the model that is of interest to us is the $(3,5)$ model,
but first we shall look at the $(3,4)$ model. In the bosonic system,
this is in fact the unitary model of the $(3,*)$ series, and is the
critical Ising model. Its string equations (in the other formalism)
were derived and studied in
refs.\cite{Brezin:1990db,Gross:1990ni,Crnkovic:1990tn}. The
non--perturbative physics of the closed string equations formulated as
reviewed in section~\ref{proposal} was studied in
refs.\cite{Johnson:1992uy,Johnson:thesis}.

In this new context, this model does not exist in its own right, and
will be a subsector of the unitary model of interest, the (3,5), which
is in fact the tricritical Ising model coupled to supergravity. We
will be able to switch on an operator in the (3,5) to approach this
(3,4) model by RG flow.  We start with the (3,4) since it is a
familiar example, and it will help us check a crucial normalisation we
will need later. The basic model has fixed $t_{1,2}=3/7$ and all other
higher $t_{l,k}=0$, and:
\begin{eqnarray}
&&{\cal R}^2=R^2_{1,2}+\frac{5}{3}t_{2,1}R^2_{2,1}+\frac{4}{3}t_{1,1}R^2_{1,1}+t_{1,0}\nonumber\\
&&{\cal R}^3=R^3_{1,2}+\frac{5}{3}t_{2,1}R^3_{2,1}+\frac{4}{3}t_{1,1}R^3_{1,1}+2t_{2,0}\ .
  \label{eq:theRs}
\end{eqnarray}
If this were a viable model in its own right (it is not in this
context), we can check the scalings of the operator content by working
directly at the level of the sphere (tree level). One can drop all
derivatives in this limit, and recall that in perturbation theory that
${\cal R}^2=0$ and ${\cal R}^3=0$. Since at this level we have:
\begin{equation}
R^2_{1,2}\sim u_2^3\sim u_3^2\ , R^2_{2,1}\sim u_2^2\ , R^2_{1,1}\sim
u_3\ , R^3_{1,2}\sim u_2u_3\ , R^3_{2,1}\sim u_3\ , R^3_{1,1}\sim u_2\ ,
  \label{eq:spherelevel}
\end{equation}
we can determine the~$z$ scaling of all of the operators:
\begin{equation}
-t_{1,0}=z\sim z^1\ ,\qquad {\cal B}=-2t_{2,0}\sim z^{\frac{5}{6}}\ ,\qquad
 t_{2,1}\sim z^{\frac13}\ .
  \label{eq:operatorscalings}
\end{equation}
Note that $t_{1,1}\sim z^{1/2}$ and can be absorbed into a boundary
operator.  The numbers in equation~\reef{eq:operatorscalings} are
indeed the dressed dimensions relative to the cosmological constant
$z\sim\mu$ in the model. Recall that this is all in the context of the
{\it bosonic} superconformal series, and one can confirm the numbers
by comparing to the continuum result given in
equation~\reef{eq:liouvilledimension}.

In the present context these numbers are irrelevant. The operator
dimensions will be recomputed in the next subsection once the model is
subsumed into the (3,5) model, which in the context of the
superconformal series, is indeed the unitary model. 

Next, we check a normalisation. The basic equations in perturbation
theory, for the case $\Gamma=0$, can be chosen to be ${\cal
  R}^2=0={\cal R}^3$, where:
\begin{equation}
{\cal
  R}^2=-\frac{1}{12}u_2^{''''}-\frac{3}{4}u_2u_2^{''}-\frac{3}{8}(u_2^\prime)^2-\frac{1}{2}u_2^3+\frac{2}{3}u_3^2-z\ ;\qquad {\cal R}^3=\quad R^3_{1,2} =\frac{2}{3}\left(u_3^{\prime\prime}+3u_3u_2\right)-{\cal B} \ ,
  \label{eq:fullRising}
\end{equation}
On the sphere, we have:
 \begin{equation}
-\frac{1}{2}u_2^3+\frac{2}{3}u_3^2=-z\ ,\qquad 2u_3u_2={\cal B}\ ,
   \label{eq:sphereising}
 \end{equation}
and we may solve the second equation, and substitute into the first,
to get
\begin{equation}
u_3=\frac{\cal B}{2u_2}\ , \qquad 3u_2^5-{\cal B}^2=6zu_2^2\ .
  \label{eq:sphereisingmore}
\end{equation}
The second polynomial equation can solved for $u_2(z,{\cal B})$, and
then using this the first equation then yields $u_3(z,{\cal B})$. A
simple solution arises when ${\cal B}=0$, for which it is seen that
$u_3=0$. This is appropriate, since ${\cal B}$ is actually the
magnetic field, and $u_3$ is the response to it, {\it i.e.,} the
magnetisation. 

Staying in  the sector where $u_3=0$ and ${\cal B}=0$, and keeping
$\Gamma=0$, we get (writing  $u=-u_2$):
\begin{equation}
{\cal
  R}^2=\frac{1}{12}u^{''''}-\frac{3}{4}uu^{\prime\prime}
-\frac{3}{6}(u^\prime)^2
  +\frac{1}{2}u^3-z\ ;\qquad {\cal R}^3=0\ ,
  \label{eq:reducedRising}
\end{equation}
(where the last equation is an identity),
 and the vanishing of ${\cal R}^2$ will yield perturbation
theory for the function $u(z)$. After a few lines of algebra, the
first few terms of the solution are:
\begin{equation}
u(z)=-2^{\frac{1}{3}}z^{\frac{1}{3}}-\frac{\nu^2}{12z^2}+\cdots\ ,
  \label{eq:perturb}
\end{equation}
and so if we use equation~\reef{partfun}, we reproduce the standard
result that the torus partition function for the gravitating Ising
model is:
\begin{equation}
Z=-\frac{1}{24}\frac{(p-1)(q-1)}{(p+q-1)}\log z \to -\frac{1}{24}\log z\ .
  \label{eq:partfunagain}
\end{equation}
This sets our normalisation for the model of interest in the next
subsection.

Let us now work with $\Gamma\neq0$ to see what we might learn. The
expansion above in equation~\reef{eq:perturb} can be taken for either
large positive~$z$ or negative~$z$. Interestingly, switching on
$\Gamma$ breaks this symmetry, in that the large positive~$z$
expansion picks up terms linear in $\Gamma$ which are imaginary. So if
we wish for $u(z)$ to be real, and to smoothly connect to non--zero
$\Gamma$ (which seems reasonable) then we are forced to pick the
expansion as being for the negative~$z$ direction. The fact that the
other choice does not smoothly connect onto non--zero~$\Gamma$, while
keeping $u(z)$ real, may well be an earmark of a branch of new
physics\footnote{We could also continue $\Gamma$ to $i\Gamma$, which
  is interesting, but we shall keep a definite choice of $\Gamma$ in
  order to compare each member of the 3--series with each other.}, but
we will not explore that here.

So, for $\Gamma\neq0$, expanding in large negative~$z$, we find
\begin{equation}
u(z)=2^{\frac{1}{3}}z^{\frac{1}{3}}-\frac{2^{2\over3}\nu\Gamma}{z^{5\over6}}
+\frac{1}{12}\frac{\nu^2}{z^2}  \left(4\Gamma^2-1\right)
 -\frac{91}{324}\frac{2^{1\over3}\nu^3}{z^{19\over6}}\Gamma\left(\Gamma^2-1\right)+\cdots\ ,
  \label{eq:perturbmore}
\end{equation}
while expanding for large positive~$z$ we find
\begin{eqnarray}
u(z)&=&-\frac{2}{3}\frac{\nu^2}{z^2}\left(3\Gamma^2+1\right)\nonumber\\
&&+{\frac {16}{81}}\frac{{\nu}^{8} }{{z}^{9}}\,\left( 3\,{{\Gamma}}^{2}+16
 \right)  \left( 3\,{{\Gamma}}^{2}+49 \right)  \left(
 3\,{{\Gamma}}^{2}+4 \right)  \left( 3\,{{\Gamma}}^{2}+
 1\right)\times\nonumber\\
&&\hskip2.5cm\Biggl\{1-\frac{10}{9}\frac{\nu^6}{z^7}
\left( 27\,{{\Gamma}}^{6}+2079\,{{\Gamma}}^{4}+44982\,{{\Gamma}}^{2}+246424 \right)\Biggr\} +\cdots\ 
  \label{eq:perturbmoremore}
\end{eqnarray}
Based on our experience with these equations, it is entirely
reasonable to expect the string equation supplies the means to
uniquely connect these two expansions smoothly through the
non--perturbative region. It would be interesting to prove this
analytically.

Integrating up twice to get the free energy, we see that the natural
dimensionless world--sheet expansion parameter ({\it i.e.}, the string
coupling) is $\kappa=\nu/z^{7/6}$. The topological expansion is very
interesting. For large negative~$z$, we see that the first two terms
in equation~\reef{eq:perturbmore} represent the sphere and the disc
respectively, while the next term contains the torus and the cylinder,
and the next the holey torus, and so forth. The interpretation is that
there are $\Gamma$ D--branes (in some units), and there is an
appropriate factor of $\Gamma^b$ for every boundary, accompanying the
usual $\kappa^{2h-2+b}$ in string perturbation theory.

For large positive~$z$, things are rather different. With the same
identification of expansion parameters as before, the presence of only
even powers of $\Gamma$ presents a puzzle. The first non--vanishing
physics is at the level of the torus and the cylinder, but then
nothing until higher orders. These orders are the four--torus (sphere
with four handles), holey sphere (it has six holes), {\it etc.}  There
are no terms like the disc, {\it etc.,} with odd numbers of
boundaries. A simpler explanation, following
ref.\cite{Klebanov:2003wg}, is to not attribute a brane
interpretation to this regime at all. Rather, there are all the
correct surfaces present for purely {\it closed string} perturbation
theory, and instead, each power of $\Gamma^2$ represents an insertion
of R--R flux. It would be interesting to examine an underlying matrix
model (such as the one in ref.\cite{Houart:1993nq}) further to add
evidence for this claim. In the present case, we shall take this as
the simplest interpretation. We could invoke other more complicated
reasons for the vanishing of the surfaces with odd numbers of
boundaries, but none have occurred to us which seem as natural as this
one.
 
With this interpretation, it is tempting to declare that we have found
our first new example of a transition between closed string physics
with fluxes and physics with D--branes present. Bear in mind however
that this (3,4) model is not a complete 0A model in its own right,
since it is not a member of the superconformal series (since 3 is odd
but 4 is not).  However, it will be (with appropriate coupling to an
operator) naturally embedded within the prototype complete 0A model we
study in the next subsection, and so the remarks here about R--R flux,
and the possibility of a non--perturbative geometric transition from
flux to D--branes, will become an honest interpretation within that
context.

\subsection{The (3,5)  model: Superdressed Tricritical Ising}
It is quite amusing that the first complete non--trivial model of
interest to us is the tricritical Ising model. This model is
fascinating in its own right since in ordinary conformal field theory
({\it i.e.,} no dressing) it is the unique minimal model which is both
conformal and superconformal~\cite{Friedan:1985rv}. In the bosonic
context, it appears gravitationally dressed as the $(4,5)$ model. In
this context, it is supergravitationally dressed and is the $(3,5)$.

The model has fixed $t_{2,2}=3/8$ and all other higher $t_{l,k}=0$,
and:
\begin{eqnarray}
&&{\cal R}^2=R^2_{2,2}+\frac{7}{3}t_{1,2}R^2_{1,2}+\frac{5}{3}t_{2,1}R^2_{2,1}+\frac{4}{3}t_{1,1}R^2_{1,1}+t_{1,0}\nonumber\\
&&{\cal R}^3=R^3_{2,2}+\frac{7}{3}t_{1,2}R^3_{1,2}+\frac{5}{3}t_{2,1}R^3_{2,1}+\frac{4}{3}t_{1,1}R^3_{1,1}+2t_{2,0}\ .
  \label{eq:moreoftheRs}
\end{eqnarray}
To compute the operator dimensions, we note that in addition to the
sphere level behaviour listed in equation~\reef{eq:spherelevel}, we have:
\begin{equation}
R^2_{2,2}\sim u_2^2u_3 \, \qquad R^3_{2,2}\sim u_2^3\sim u_3^2\ ,
 \label{eq:spherelevelmore}
\end{equation}
From the second equation we see that $u_2\sim {\cal B}^{1/3}$ and
$u_3\sim {\cal B}^{1/2}$, and so a little algebra gives the~$z$
scaling of all of the operators:
\begin{equation}
-t_{1,0}=z\sim z^1\ ,\qquad {\cal B}=-2t_{2,0}\sim z^{\frac{6}{7}}\ ,\qquad
 t_{1,2}\sim z^{\frac17}\ ,\qquad t_{1,1}\sim z^{\frac{4}{7}}\ .
  \label{eq:operatorscalingsmore}
\end{equation}
Note that $t_{2,1}\sim z^{3/7}$ and can be absorbed into a boundary
operator. The operator scalings in
equation~\reef{eq:operatorscalingsmore} are precisely the scalings
required by the formula~\reef{eq:superliouvilledimension2} (derived
with continuum methods) to work, and the model is to be interpreted as
a type~0A background, as per our proposal.

Again, let us study perturbation theory with $\Gamma=0$ first. Then we
have a choice ${\cal R}^3=0={\cal R}^2$ where now (switching off
$t_{1,1}$):
\begin{eqnarray}
&&{\cal R}^2=-\frac{1}{9}u_3^{''''}-\frac{5}{6}u_2u_3^{\prime\prime}-\frac{5}{12}u_2^{\prime\prime}u_3-\frac{5}{12}u_2^\prime
u_3^\prime-\frac{5}{4}u_2^2u_3 -z\ ,\nonumber\\
&&{\cal R}^3=-\frac{1}{6}u_2^{''''}-\frac{5}{4}u_2u_2^{\prime\prime}-\frac{15}{16}(u_2^\prime)^2 -\frac{5}{8}u_2^3+\frac{5}{3}u_3^2 -{\cal B} \ ,
  \label{eq:fullRtricritical}
\end{eqnarray}
and so on the sphere we have the equations
\begin{equation}
-\frac{5}{4}u_2^2u_3 =z\ ,\qquad \frac{5}{3}u_3^2 ={\cal B} \ .
  \label{eq:tricriticalisingsphere}
\end{equation}
We may solve the first equation and substitute into the second to get:
\begin{equation}
u_3=-\frac{4}{5}\frac{z}{u_2^2}\ ,\qquad \frac{5}{8}u_2^7+{\cal
  B}u_2^4=\frac{16}{15}z^2\ .
  \label{eq:tricriticalisingspheremore}
\end{equation}
Again, this system can be used to solve for $u_2(z,{\cal B})$ and
$u_3(z,{\cal B})$. Notice that this time (in contrast to the case in
the previous section) we can choose ${\cal B}=0$ and still retain
non--zero $u_3$. In fact, after a bit of algebra, we find for this
choice, writing ($u_2=-u$):
\begin{equation}
u=-\frac{2}{75}(75)^{\frac{6}{7}}z^{\frac27}\ ,\qquad
u_3=-\frac{1}{5}(75)^{\frac27}z^{\frac67}\ .
  \label{eq:tricriticalsphere}
\end{equation}
Let us now go beyond tree level to uncover a
bit more physics, staying with $\Gamma=0$ for now. More algebra yields
from our equations the next few levels of closed string perturbation
theory:
\begin{eqnarray}
&&
u=-\frac{2}{75}(75)^{\frac{6}{7}}z^{\frac27}-{2\over 1029}{\nu^2\over
  z^2}+\frac{240242}{5294205}(75)^{\frac17}{\nu^4\over
  z^{\frac{30}{7}}}+\cdots\ ,\nonumber\\
&&
u_3=-\frac{1}{5}(75)^{\frac27}z^{\frac67}+\frac{24}{1715}75^{\frac{3}{7}}\frac{\nu^2}{z^{\frac{13}{7}}}+
  {116003\over 8823675}(75)^{\frac47}{\nu^4\over z^{\frac{29}{7}}}+\cdots\ .
  \label{eq:tricriticalperturbed}
\end{eqnarray}
We can readily integrate the above result for $u$ twice to get the
expression for the free energy, using equation~\reef{partfun}.
We see that the dimensionless world sheet expansion parameter (string
coupling) is $\kappa=\nu/z^{8/7}$.

Now we switch $\Gamma$ back on, and solve the full
string equation~\reef{eq:thestringequation} perturbatively, starting
with the same sphere level solutions.  We find that the next
non--vanishing level of perturbation theory is of order $\nu$, and
denoting the contributions to $u$ and $u_3$ at this order by $f$ and
$g$, respectively, we have the following 
equation for them:
\begin{equation}
f(z)=\frac{(75)^{\frac67}}{7875z^{14}}\left\{
  -7(75)^{\frac57}z^{97\over 7}g(z)\pm
  5\cdot 7^{1\over7}\sqrt{(75)^{1\over7}z^{\frac{184}{7}}\left(49
      (75)^{2\over7}z^{10\over7}g(z)^2+36\Gamma^2\right)}\right\}\ .
  \label{eq:mustsolve}
\end{equation}
This dreadful equation can be simplified by realising that the
solution for $g(z)$ must reduce to zero at $\Gamma=0$, and similarly
for $f(z)$, since we would like to match onto closed string
perturbation theory in this limit. One way that we can achieve this is
by setting the square root to zero, since $g(z)\propto\Gamma$ for this
choice, and we remark in passing that other choices of $g(z)$ will
given two other branches of solutions. It would be interesting to
explore these further. In proceeding to develop the non--zero $\Gamma$
perturbation theory, we note that (rather like happened in the case of
the previous subsection) in order to have real solutions for $u$ and
$u_3$, we must take our expansion as pertaining to negative $z$.
Carrying on with the choice in equation~\reef{eq:mustsolve} that makes
the square root vanish, after going on to the next few levels of
perturbation theory we have:
\begin{eqnarray}
&&
u=-\frac{2}{75}(75)^{\frac{6}{7}}z^{\frac27}\mp\frac{2}{35}(75)^{\frac37}\frac{\nu\Gamma}{z^{\frac47}}-{2\over 1029}{\nu^2\over
  z^2}+0\,\,\nu^3+\frac{240242}{5294205}(75)^{\frac17}{\nu^4\over
  z^{\frac{30}{7}}}
+\cdots\ ,\nonumber\\
&&
u_3=-\frac{1}{5}(75)^{\frac27}z^{\frac67}\pm\frac{6}{7}(75)^{-\frac17}\frac{\nu\Gamma}{z^{\frac57}}+
\frac{3}{1715}(75)^{\frac{3}{7}}\frac{\nu^2(21\Gamma^2+8)}{z^{\frac{13}{7}}}+\frac{52}{2401}\frac{\nu^3\Gamma}{z^3}\nonumber\\
&&\hskip5cm - {(75)^{\frac47}\over 8823675}{\nu^4\over z^{\frac{29}{7}}}\left(33075\Gamma^4-300720\Gamma^2-116003\right)
+\cdots\ 
  \label{eq:disclevel}
\end{eqnarray}
It is very curious that there is no contribution to $u$ of surfaces
with boundary beyond the disc, although this is not the case for
$u_3$. Similarly interesting is the choice to make $g(z)$ vanish in
equation~\reef{eq:mustsolve}. In that case, it is the $u_3$
perturbative series which fails to be afflicted by $\Gamma$ at higher
orders, while the $u$ perturbative series has contributions from all
orders. However, it is clear that for generic choices of $g(z)$, we
will have contributions to the physics at all orders in perturbation
theory from all types of surfaces with and without boundaries.

The reader should not be concerned that these choices for $g(z)$
correspond to non--perturbative ambiguities in the physics. There is
nothing about the properties of the string equation which suggest that
this is the case, as far as we know. It must be recalled that there
are also constraints that cannot be specified in the perturbative
analysis, such as the nature of the asymptotic expansion in the other
large $z$ regime. Generically, we expect that the requirement of
fixing to perturbation theory in {\it both} the $z\to+\infty$ and the
$z\to-\infty$ regimes will fix things uniquely. The present analysis
is simply a demonstration of the perturbative possibilities available
to us.

Turning to a wider range of choices, note that we can start at
sphere level with vanishing $u=-u_2$. This is now physics that the
equations ${\cal R}^3=0={\cal R}^2$ cannot handle, but the full string
equation can, since the sphere level condition is:
\begin{equation}
\frac{1}{2}u_2({\cal R}^2)^2+u_3{\cal R}^2{\cal
  R}^3-\frac{1}{4}u_2^2({\cal R}^3)^2=0\ ,
  \label{eq:spherecondition}
\end{equation}
where
\begin{equation}
{\cal R}^2=-\frac{5}{4}u_2^2u_3 -z \quad{\rm  and}\quad 
{\cal R}^3= \frac{5}{3}u_3^2 {\cal B} \ .
  \label{eq:Rdefssphere}
\end{equation}
The choice $u_2=0$ leaves:
\begin{equation}
u_3=\pm\frac{(15)^{\frac12}}{5}{\cal B}^{\frac12}\ .
  \label{eq:youthreechoice}
\end{equation}
We can choose ${\cal B}$ to vanish, as a special case, resulting in
vanishing $u_3$ on the sphere. There are no doubt solutions for which
$u_3$ is non--zero beyond tree level, but we won't pursue those here.

It is also interesting to consider the case of $u_3$ vanishing to all
orders and beyond. In such a case, we have in the string equation the
functions:
\begin{equation}
 {\cal
   R}^3=\frac{1}{6}u^{''''}-\frac{5}{4}uu^{\prime\prime}
 -\frac{15}{16}(u^\prime)^2
   +\frac{5}{8}u^3\ ;\qquad {\cal R}^2=-z\ .
   \label{eq:reducedR}
\end{equation}

Expanding, we have for example in the
$z\to+\infty$ limit:
\begin{eqnarray}
 \hskip-0.5cm u(z)&=&-\frac{2}{3}
{\frac {{\nu}^{2} \left( 3\,{{\Gamma}}^{2}+1 \right) }{{z}^{2
}}}\nonumber\\
&&
-{\frac {50}{6561}}\,\frac{{\nu}^{16}}{{z}^{18}} \left( 3\,{{\Gamma}}^{2}+1 \right) ^{2} \left( 3\,{{
\Gamma}}^{2}+4 \right)^2 \left( 3\,{{\Gamma}}^{2}+25
 \right)^2\left(3\Gamma^2+64\right)\left(3\Gamma^2+16\right)\times\nonumber\\
&&\Biggl\{1+\frac{100}{2187}\frac{\nu^{14}}{z^{16}} 
\left( 3\,{{\Gamma}}^{2}+1 \right) \left( 3\,{{
\Gamma}}^{2}+4 \right)\left( 3\,{{\Gamma}}^{2}+25
 \right) \times\nonumber\\
&&\hskip1.3cm\left(162\Gamma^8  + 41931\Gamma^6  + 4316193\Gamma^4  + 187005426\Gamma^2  + 2729467376\right)
\Biggr\}+\cdots
  \label{eq:asymptoticB}
\end{eqnarray}
The physics of this regime is again expected to be connected
by smooth non--perturbative physics through the strong coupling region
to other perturbative physics, such as that given in
equation~\reef{eq:tricriticalperturbed}, and it would be interesting
to pursue this further.

Notice that we can also have a solution for which $u\sim z^{\frac27}$
and $u_3=0$ at tree level, in the large negative $z$ regime, which is
again something which is allowed by the sphere
condition~\reef{eq:spherecondition} of the full string equation, and
not the restriction to the solutions of ${\cal R}^3=0={\cal R}^2$. The
first few terms of the asymptotic expansion of the solution of our
string equation are:
 \begin{eqnarray}
 \hskip-0.5cm u(z)&=&-\frac{2}{5}{5}^{\frac{5}{7}}{z}^{\frac{2}{7}}
 +{\frac {2{\nu}^{2} }{1029}}\,{\frac {\left( 147\,{{\Gamma}}^{2} -43\right) }{{z}^{2}}}+\frac {4\cdot{5}^{\frac{2}{7}}  }{1058841}\frac{{\nu}^{4}}{z^{{\frac {30}{7}}}} \, \left( 43218 \Gamma^4  - 234318 \Gamma^2 +15389 \right)
 \nonumber\\
 &&+\frac {8\cdot 5^{\frac47}  }{363182463 }\frac{\nu^6 }{z^{{\frac
 {46}{7}}}}
\left(  6353046 \Gamma^6 - 104047335 \Gamma^4  + 326118177 \Gamma^2  +  5183425 \right) 
 +\cdots\ 
   \label{eq:asymptotic}
 \end{eqnarray}
 
 Examining this expression and the previous one, tentatively
 associating each power of $\Gamma$ with a boundary, we have the
 sphere, torus and cylinder from the first two terms, but no disc. The
 next term contains the double torus, torus with two holes and sphere
 with four holes. Missing are any surfaces with an odd number of
 holes.  Following on from what we have already stated about the
 expansions observed in the previous subsections, (and based on
 examples in ref.\cite{Klebanov:2003wg}), this is highly suggestive of
 a R--R flux interpretation instead of boundaries. So tentatively,
 there are no world sheet boundaries in the expansion but fluxes: one
 insertion for every power of $\Gamma^2$, however this needs to be
 explored more.
 
 It is our conjecture that it will {\it always} be the case that only
 even powers of $\Gamma$ appear in solutions for which the behaviour
 on the sphere is a solution of equation~\reef{eq:spherecondition} but
 not ${\cal R}^3=0={\cal R}^2$, and further, that it extends to all of
 the string equations for the $(q,p)$ series. A proof of this would be
 useful to construct.
 
 So we have a number of possibilities, depending upon the boundary
 conditions. We remind the reader that the overall situation is
 necessarily more complicated that for the study of the
 equation~\reef{nonperturbative}, for the $(2,4k)$ series because
 there are two functions to solve for, $u_2$ and $u_3$, in addition to
 a number of couplings, ${\cal B}$ and $t_{1,2}$. It is clear that we
 can have solutions which have D--branes in the large negative~$z$
 regime, but only fluxes for positive~$z$. These are new fully
 non--perturbative examples of geometrical transitions. We may well
 also have solutions which have fluxes in {\it both} regime, which is
 certainly novel. These would arise if we choose sphere behaviour in
 both asymptotic regions which are allowed by the tree level
 condition~\reef{eq:spherecondition} of the full string equation, but
 not ${\cal R}^3=0={\cal R}^2$. We do not know if this is consistent
 non--perturbatively, but certainly seems like a possibility at the
 level of our current analysis. 
 
 For either the former or the latter case, we can always expect
 transition phenomena for another reason: We must not forget that we
 have switched off a number of operators in the theory.  Recall from
 the explorations in the last subsection of the (3,4) model, that the
 large negative~$z$ region contained branes, while the large
 positive~$z$ direction contained only fluxes. There was a geometric
 transition as one went through the strong coupling regime.  If we
 switch on the operator ${\cal O}_{1,2}$ by turning on $t_{1,2}$ it
 will mix the physics we have seen here with the physics of the (3,4)
 model. So in fact, we have a new way of generating transition
 phenomena, by turning on ${\cal O}_{1,2}$, even if we started with a
 solution in the (3,5) model with no transition.
 
 Let us see how this must work, continuing to keep $u_3=0$ and ${\cal
   B}=0$. We must now solve the string equation with:
\begin{equation}
{\cal
  R}^3=\frac{1}{6}u^{''''}-\frac{5}{4}uu^{\prime\prime}
-\frac{15}{16}(u^\prime)^2
  +\frac{5}{8}u^3\ ;\qquad {\cal R}^2=\frac{7}{3}t_{1,2}\left(\frac{1}{12}u^{''''}-\frac{3}{4}uu^{\prime\prime}
-\frac{3}{8}(u^\prime)^2
  +\frac{1}{2}u^3\right)-z\ .
  \label{eq:reducedRless}
\end{equation}
Unfortunately, this becomes a rather difficult problem to proceed with
analytically. Already at the sphere, after working to linear order in
$t_{1,2}$, we see that the physics is controlled by the solution to
this equation:
\begin{equation}
0=-\frac{25}{256}u^7-\frac{1}{2} z^2+ \frac{7}{3}zt_{1,2} \left( \frac{1}{2}u^3- z \right)\ .
  \label{eq:mixedsphere}
\end{equation}
The next step will be to solve this for $u$ at linear order in
$t_{1,2}$ and then proceed. It is not clear if the mixing with new
world sheets with boundaries will ever be seen in the linear
approximation however, or even to any order in perturbation in
$t_{1,2}$, so this approach may be doomed. However, we do know that
ultimately, since $t_{1,2}$ is a good operator in the theory there
will be a flow to the behaviour of the previous subsection givign a
new handle with which to switch on transition phenomena. 

It is also possible that switching on  the other coupling which we
have set to zero (${\cal B}$), together with the function $u_3(z)$ for
large positive~$z$, may also allow for odd powers of $\Gamma$ to
appear, giving us even more directions in which we can move to find
transitions. It is also likely that there is an interesting
geometrical description of this moduli space of directions giving a
nicer setting in which some of these transition phenomena might have a
powerful description. The presence of an underlying supersymmetric
model makes this seem worth seeking.

Further study of this $(3,5)$ example ought to be carried out.  We've
looked at a range of perturbative solutions, and showed that examples
of the interesting physics which we seek abound, but further analysis
of how these asymptotes connect to each other non--perturbatively
ought to be performed by studying the string equation further.

\subsection{The Torus Partition Function}

Finally, we note two more interesting pieces of data, the one--loop
contribution to the partition function. This should give useful
universal data which can be compared to the continuum theory. For
large $-z$ regime the torus contribution is:
\begin{equation}
Z_A(z)=-\frac{1}{1029}\log z\ ,
  \label{eq:torusA}
\end{equation}
while in the $+z$ regime it is:
\begin{equation}
Z_A(z)=-\frac{1}{3}\log z\ .
  \label{eq:torusB}
\end{equation}

The latter result is particularly interesting. First of all, it
appears to be a universal contribution yielded by all solutions in the
$(3,p)$ series which begin with sphere contributions not satisfying
${\cal R}^3={0}={\cal R}^2$, but satisfying~\reef{eq:spherecondition}.
The expected full result\cite{Bershadsky:1991zs}:
\begin{equation}
Z_{\rm even}=\frac{1}{2}\left(Z_A(z)+Z_B(z)\right)=
-\frac{1}{8}\frac{(p-1)}{(p+1)}\log z\ ,\qquad p\,\,
{\rm odd}\ ,\,\, p\neq0 \,\,{\rm mod}\,\,3\ ,
  \label{eq:spinstructure}
\end{equation}
would seem to require that the 0B contribution in this regime is
non--vanishing.  (Interestingly, it vanishes exactly for the $(2,4k)$
series\cite{Klebanov:2003wg}, but this had better not happen here
since the type~0B contribution is needed to restore the $p$ dependence
of the formula. If our proposal is correct, this number serves as a
prediction for the type~0B definition (yet to be presented), in order
to make contact with the continuum results.

\section{Closing Remarks}

Here, we shall be brief, since we have made the key technical
observations along the way.  It is hoped that it is apparent to the
reader that the old double scaled large $N$ matrix technology,
although failing to tell us anything directly about higher dimensional
string theories, may still supply useful laboratories within which to
study a range of phenomena of current interest. This should be
definitely appreciated once it is realised that there are very many
well--defined non--perturbative formulations of the physics. These
models remain the only real examples we have in string theory where
certain questions can be asked and answered at arbitrary values of the
coupling. Admittedly, given that the string theories which can be
formulated with these methods do not admit a lot of room for a wide
variety of physics, the range of questions is rather smaller than for
higher dimensional string theory, but there is still scope for lessons
to be learned.

Here, we have proposed a natural formulation of the type 0A string in
${\hat c}\leq 1$ backgrounds, the superconformal minimal models
coupled to supergravity, extending the work of
ref.\cite{Klebanov:2003wg}. This work is all based on
non--perturbative formulations of string theory, including both closed
and open sectors, which have existed in the literature for over a
decade now, but have been recast in a new role as type~0A theories.
The first non--trivial new model which we studied in this paper, the
(supergravitationally dressed) tricritical Ising model as it turns
out, already supplied us with new physics which should be studied
further, and presumably there is more to be found in other examples
too.  There is a lot more to learn about these models, and no doubt the
remarkably rich underlying structures (integrable hierarchies,
W--algebra constraints and so forth) may play an important role in
this new context.

\section*{Acknowledgements}CVJ thanks Juan Maldacena and Nathan
Seiberg for comments.  CVJ's research while at the ITP was
supported in part by National Science Foundation under Grant No.
PHY-99-07949.


\begin{thebibliography}{10}

\bibitem{Shenker:1990uf}
S.~H. Shenker, ``The Strength of nonperturbative effects in string theory,''.
  Presented at the Cargese Workshop on Random Surfaces, Quantum Gravity and
  Strings, Cargese, France, May 28 - Jun 1, 1990.

\bibitem{Johnson:1998pc}
C.~V. Johnson, ``Etudes on D-branes,''
\href{http://www.arXiv.org/abs/hep-th/9812196}{{\tt hep-th/9812196}}.

\bibitem{McGreevy:2003kb}
J.~McGreevy and H.~Verlinde, ``Strings from tachyons: The c = 1 matrix
  reloated,''
\href{http://www.arXiv.org/abs/hep-th/0304224}{{\tt hep-th/0304224}}.

\bibitem{Martinec:2003ka}
E.~J. Martinec, ``The annular report on non-critical string theory,''
\href{http://www.arXiv.org/abs/hep-th/0305148}{{\tt hep-th/0305148}}.

\bibitem{Klebanov:2003km}
I.~R. Klebanov, J.~Maldacena, and N.~Seiberg, ``D-brane decay in
  two-dimensional string theory,'' {\em JHEP} {\bf 07} (2003) 045,
\href{http://www.arXiv.org/abs/hep-th/0305159}{{\tt hep-th/0305159}}.

\bibitem{McGreevy:2003ep}
J.~McGreevy, J.~Teschner, and H.~Verlinde, ``Classical and quantum D-branes in
  2D string theory,''
\href{http://www.arXiv.org/abs/hep-th/0305194}{{\tt hep-th/0305194}}.

\bibitem{Alexandrov:2003nn}
S.~Y. Alexandrov, V.~A. Kazakov, and D.~Kutasov, ``Non-perturbative effects in
  matrix models and D-branes,'' {\em JHEP} {\bf 09} (2003) 057,
\href{http://www.arXiv.org/abs/hep-th/0306177}{{\tt hep-th/0306177}}.

\bibitem{Fateev:2000ik}
V.~Fateev, A.~B. Zamolodchikov, and A.~B. Zamolodchikov, ``Boundary Liouville
  field theory. I: Boundary state and boundary two-point function,''
\href{http://www.arXiv.org/abs/hep-th/0001012}{{\tt hep-th/0001012}}.

\bibitem{Schomerus:2003vv}
V.~Schomerus, ``Rolling tachyons from Liouville theory,''
\href{http://www.arXiv.org/abs/hep-th/0306026}{{\tt hep-th/0306026}}.

\bibitem{Sen:2003iv}
A.~Sen, ``Open-closed duality: Lessons from matrix model,''
\href{http://www.arXiv.org/abs/hep-th/0308068}{{\tt hep-th/0308068}}.

\bibitem{Sen:1999mg}
A.~Sen, ``Non-BPS states and branes in string theory,''
\href{http://www.arXiv.org/abs/hep-th/9904207}{{\tt hep-th/9904207}}.

\bibitem{Sen:2002nu}
A.~Sen, ``Rolling tachyon,'' {\em JHEP} {\bf 04} (2002) 048,
\href{http://www.arXiv.org/abs/hep-th/0203211}{{\tt hep-th/0203211}}.

\bibitem{Dalley:1992qg}
S.~Dalley, C.~V. Johnson, and T.~Morris, ``Multicritical complex matrix models
  and nonperturbative 2-D quantum gravity,'' {\em Nucl. Phys.} {\bf B368}
  (1992)
625--654.

\bibitem{Dalley:1992vr}
S.~Dalley, C.~V. Johnson, and T.~Morris, ``Nonperturbative two-dimensional
  quantum gravity,'' {\em Nucl. Phys.} {\bf B368} (1992)
655--670.

\bibitem{Dalley:1992yi}
S.~Dalley, C.~V. Johnson, and T.~Morris, ``Nonperturbative two-dimensional
  quantum gravity, again,'' {\em Nucl. Phys. Proc. Suppl.} {\bf 25A} (1992)
  87--91,
\href{http://www.arXiv.org/abs/hep-th/9108016}{{\tt hep-th/9108016}}.

\bibitem{Moore:1992sf}
G.~W. Moore, ``Double scaled field theory at c = 1,'' {\em Nucl. Phys.} {\bf
  B368} (1992)
557--590.

\bibitem{Moore:1992zv}
G.~W. Moore, M.~R. Plesser, and S.~Ramgoolam, ``Exact S matrix for 2-D string
  theory,'' {\em Nucl. Phys.} {\bf B377} (1992) 143--190,
\href{http://www.arXiv.org/abs/hep-th/9111035}{{\tt hep-th/9111035}}.

\bibitem{Dhar:1995gw}
A.~Dhar, G.~Mandal, and S.~R. Wadia, ``Discrete state moduli of string theory
  from the C=1 matrix model,'' {\em Nucl. Phys.} {\bf B454} (1995) 541--560,
\href{http://www.arXiv.org/abs/hep-th/9507041}{{\tt hep-th/9507041}}.

\bibitem{Takayanagi:2003sm}
T.~Takayanagi and N.~Toumbas, ``A matrix model dual of type 0B string theory in
  two dimensions,'' {\em JHEP} {\bf 07} (2003) 064,
\href{http://www.arXiv.org/abs/hep-th/0307083}{{\tt hep-th/0307083}}.

\bibitem{Douglas:2003up}
M.~R. Douglas, I.~R. Klebanov, D.~Kutasov, J.~Maldacena, E.~Martinec, and
  N.~Seiberg, ``A new hat for the c = 1 matrix model,''
\href{http://www.arXiv.org/abs/hep-th/0307195}{{\tt hep-th/0307195}}.

\bibitem{Brezin:1990rb}
E.~Brezin and V.~A. Kazakov, ``\MakeLowercase{EXACTLY SOLVABLE FIELD THEORIES
  OF CLOSED STRINGS},'' {\em Phys. Lett.} {\bf B236} (1990)
144--150.

\bibitem{Douglas:1990ve}
M.~R. Douglas and S.~H. Shenker, ``\MakeLowercase{STRINGS IN LESS THAN
  ONE-DIMENSION},'' {\em Nucl. Phys.} {\bf B335} (1990)
635.

\bibitem{Gross:1990vs}
D.~J. Gross and A.~A. Migdal, ``\MakeLowercase{NONPERTURBATIVE TWO-DIMENSIONAL
  QUANTUM GRAVITY},'' {\em Phys. Rev. Lett.} {\bf 64} (1990)
127.

\bibitem{Gross:1990aw}
D.~J. Gross and A.~A. Migdal, ``\MakeLowercase{A NONPERTURBATIVE TREATMENT OF
  TWO-DIMENSIONAL QUANTUM GRAVITY},'' {\em Nucl. Phys.} {\bf B340} (1990)
333--365.

\bibitem{Kazakov:1990cq}
V.~A. Kazakov, ``\MakeLowercase{A SIMPLE SOLVABLE MODEL OF QUANTUM FIELD THEORY
  OF OPEN STRINGS},'' {\em Phys. Lett.} {\bf B237} (1990)
212.

\bibitem{Kostov:1990nf}
I.~K. Kostov, ``\MakeLowercase{EXACTLY SOLVABLE FIELD THEORY OF D = 0 CLOSED
  AND OPEN STRINGS},'' {\em Phys. Lett.} {\bf B238} (1990)
181.

\bibitem{Dalley:1992br}
S.~Dalley, C.~V. Johnson, T.~R. Morris, and A.~Watterstam, ``Unitary matrix
  models and 2-D quantum gravity,'' {\em Mod. Phys. Lett.} {\bf A7} (1992)
  2753--2762,
\href{http://www.arXiv.org/abs/hep-th/9206060}{{\tt hep-th/9206060}}.

\bibitem{Periwal:1990gf}
V.~Periwal and D.~Shevitz, ``\MakeLowercase{UNITARY MATRIX MODELS AS EXACTLY
  SOLVABLE STRING THEORIES},'' {\em Phys. Rev. Lett.} {\bf 64} (1990)
1326.

\bibitem{Periwal:1990qb}
V.~Periwal and D.~Shevitz, ``\MakeLowercase{EXACTLY SOLVABLE UNITARY MATRIX
  MODELS: MULTICRITICAL POTENTIALS AND CORRELATIONS},'' {\em Nucl. Phys.} {\bf
  B344} (1990)
731--746.

\bibitem{Klebanov:2003wg}
I.~R. Klebanov, J.~Maldacena, and N.~Seiberg, ``Unitary and complex matrix
  models as 1-d type 0 strings,''
\href{http://www.arXiv.org/abs/hep-th/0309168}{{\tt hep-th/0309168}}.

\bibitem{Gopakumar:1998ki}
R.~Gopakumar and C.~Vafa, ``On the gauge theory/geometry correspondence,'' {\em
  Adv. Theor. Math. Phys.} {\bf 3} (1999) 1415--1443,
\href{http://www.arXiv.org/abs/hep-th/9811131}{{\tt hep-th/9811131}}.

\bibitem{Klebanov:2000hb}
I.~R. Klebanov and M.~J. Strassler, ``Supergravity and a confining gauge
  theory: Duality cascades and $\chi$SB-resolution of naked singularities,''
  {\em JHEP} {\bf 08} (2000) 052,
\href{http://www.arXiv.org/abs/hep-th/0007191}{{\tt hep-th/0007191}}.

\bibitem{Maldacena:2000yy}
J.~M. Maldacena and C.~Nunez, ``Towards the large N limit of pure N = 1 super
  Yang Mills,'' {\em Phys. Rev. Lett.} {\bf 86} (2001) 588--591,
\href{http://www.arXiv.org/abs/hep-th/0008001}{{\tt hep-th/0008001}}.

\bibitem{Vafa:2000wi}
C.~Vafa, ``Superstrings and topological strings at large N,'' {\em J. Math.
  Phys.} {\bf 42} (2001) 2798--2817,
\href{http://www.arXiv.org/abs/hep-th/0008142}{{\tt hep-th/0008142}}.

\bibitem{Cachazo:2001jy}
F.~Cachazo, K.~A. Intriligator, and C.~Vafa, ``A large N duality via a
  geometric transition,'' {\em Nucl. Phys.} {\bf B603} (2001) 3--41,
\href{http://www.arXiv.org/abs/hep-th/0103067}{{\tt hep-th/0103067}}.

\bibitem{Morris:1990bw}
T.~R. Morris, ``2-D quantum gravity, multicritical matter and complex
  matrices,''. FERMILAB-PUB-90-136-T.

\bibitem{Morris:1992zr}
T.~R. Morris, ``Multicritical matter from complex matrices,'' {\em Class.
  Quant. Grav.} {\bf 9} (1992)
1873--1881.

\bibitem{Johnson:1992pu}
C.~V. Johnson, T.~R. Morris, and A.~Watterstam, ``Global KdV flows and stable
  2-D quantum gravity,'' {\em Phys. Lett.} {\bf B291} (1992) 11--18,
\href{http://www.arXiv.org/abs/hep-th/9205056}{{\tt hep-th/9205056}}.

\bibitem{Johnson:1992uy}
C.~V. Johnson, T.~Morris, and B.~Spence, ``Stable nonperturbative minimal
  models coupled to 2-D quantum gravity,'' {\em Nucl. Phys.} {\bf B384} (1992)
  381--410,
\href{http://www.arXiv.org/abs/hep-th/9203022}{{\tt hep-th/9203022}}.

\bibitem{Johnson:1994vk}
C.~V. Johnson, ``On integrable $c < 1$ open string theory,'' {\em Nucl. Phys.}
  {\bf B414} (1994) 239--266,
\href{http://www.arXiv.org/abs/hep-th/9301112}{{\tt hep-th/9301112}}.

\bibitem{Houart:1993nq}
L.~Houart, ``Explicit resolution of an integrable c(4,3) open string theory,''
  {\em Phys. Lett.} {\bf B311} (1993) 71--75,
\href{http://www.arXiv.org/abs/hep-th/9303157}{{\tt hep-th/9303157}}.

\bibitem{Johnson:thesis}
C.~V. Johnson, {\em {\rm ``Non--Perturbatively Stable Conformal Minimal Models
  Coupled to Two Dimensional Quantum Gravity''}}.
\newblock PhD thesis, \rm Southampton University (UK), 1992.

\bibitem{Gelfand:1976A}
I.~M. Gel'fand and L.~A. Dikii, ``Fractional Powers of Operators and
  Hamiltonian Systems,'' {\em Funct.\ Anal.\ Appl.} {\bf 10} (1976) 259.

\bibitem{Gelfand:1976B}
I.~M. Gel'fand and L.~A. Dikii, ``The Resolvent and Hamiltonian Systems,'' {\em
  Funct.\ Anal.\ Appl.} {\bf 11} (1976)
93.

\bibitem{Douglas:1990dd}
M.~R. Douglas, ``\MakeLowercase{STRINGS IN LESS THAN ONE-DIMENSION AND THE
  GENERALIZED K-D- V HIERARCHIES},'' {\em Phys. Lett.} {\bf B238} (1990)
176.

\bibitem{Gelfand:1975rn}
I.~M. Gel'fand and L.~A. Dikii, ``Asymptotic behavior of the resolvent of
  Sturm-Liouville equations and the algebra of the Korteweg-De Vries
  equations,'' {\em Russ. Math. Surveys} {\bf 30} (1975)
77--113.

\bibitem{Johnson:1992wr}
C.~V. Johnson, T.~R. Morris, and P.~L. White, ``The Boundary cosmological
  constant in stable 2-D quantum gravity,'' {\em Phys. Lett.} {\bf B292} (1992)
  283--289,
\href{http://www.arXiv.org/abs/hep-th/9206066}{{\tt hep-th/9206066}}.

\bibitem{Martinec:1991ht}
E.~J. Martinec, G.~W. Moore, and N.~Seiberg, ``Boundary operators in 2-D
  gravity,'' {\em Phys. Lett.} {\bf B263} (1991)
190--194.

\bibitem{Ginsparg:1993is}
P.~Ginsparg and G.~W. Moore, ``Lectures on 2-D gravity and 2-D string theory,''
\href{http://www.arXiv.org/abs/hep-th/9304011}{{\tt hep-th/9304011}}.

\bibitem{Brezin:1990db}
E.~Brezin, M.~R. Douglas, V.~Kazakov, and S.~H. Shenker, ``\MakeLowercase{THE
  ISING MODEL COUPLED TO 2-D GRAVITY: A NONPERTURBATIVE ANALYSIS},'' {\em Phys.
  Lett.} {\bf B237} (1990)
43.

\bibitem{Gross:1990ni}
D.~J. Gross and A.~A. Migdal, ``\MakeLowercase{NONPERTURBATIVE SOLUTION OF THE
  ISING MODEL ON A RANDOM SURFACE},'' {\em Phys. Rev. Lett.} {\bf 64} (1990)
717.

\bibitem{Crnkovic:1990tn}
C.~Crnkovic, P.~Ginsparg, and G.~W. Moore, ``\MakeLowercase{THE ISING MODEL,
  THE YANG-LEE EDGE SINGULARITY, AND 2-D QUANTUM GRAVITY},'' {\em Phys. Lett.}
  {\bf B237} (1990)
196.

\bibitem{Friedan:1985rv}
D.~Friedan, Z.-A. Qiu, and S.~H. Shenker, ``\MakeLowercase{SUPERCONFORMAL
  INVARIANCE IN TWO-DIMENSIONS AND THE TRICRITICAL ISING MODEL},'' {\em Phys.
  Lett.} {\bf B151} (1985)
37--43.

\bibitem{Bershadsky:1991zs}
M.~Bershadsky and I.~R. Klebanov, ``Partition functions and physical states in
  two-dimensional quantum gravity and supergravity,'' {\em Nucl. Phys.} {\bf
  B360} (1991)
559--585.

\end{thebibliography}

\providecommand{\href}[2]{#2}\begingroup\raggedright\endgroup

\end{document}